\documentclass[runningheads]{llncs}
\usepackage[T1]{fontenc}
\usepackage{graphicx}
\usepackage{float} 
\usepackage{xcolor}
\usepackage{comment}
\usepackage{multirow}
\usepackage[english]{babel}
\usepackage{hyperref}

\usepackage{wrapfig}

\usepackage{array}
\usepackage{booktabs}

\usepackage{amsfonts}
\usepackage{amssymb}
\usepackage{amsmath}
\usepackage{siunitx}
\usepackage[skip=10pt plus1pt]{parskip}

\usepackage{cite}

\begin{document}
\title{Machine Learning Methods as Robust Quantum Noise Estimators}
\titlerunning{Machine Learning as Quantum Noise Estimators}
%
\author{Jon Gardeazabal-Gutierrez\orcidID{0009-0008-8111-6019} \and Erik B. Terres-Escudero\orcidID{0009-0003-9781-7657}  \and Pablo García Bringas\orcidID{0000-0003-3594-9534}}

\authorrunning{Gardeazabal-Gutierrez et al.}
%
\institute{Deusto University, Bilbao, Spain \footnote{Corresponding author: \email{jon.g.g@deusto.es}}}

%
%

%
\maketitle              
\begin{abstract}

Access to quantum computing is steadily increasing each year as the speed advantage of quantum computers solidifies with the growing number of usable qubits. However, the inherent noise encountered when running these systems can lead to measurement inaccuracies, especially pronounced when dealing with large or complex circuits. Achieving a balance between the complexity of circuits and the desired degree of output accuracy is a nontrivial yet necessary task for the creation of production-ready quantum software. In this study, we demonstrate how traditional machine learning (ML) models can estimate quantum noise by analyzing circuit composition. To accomplish this, we train multiple ML models on random quantum circuits, aiming to learn to estimate the discrepancy between ideal and noisy circuit outputs. By employing various noise models from distinct IBM systems, our results illustrate how this approach can accurately predict the robustness of circuits with a low error rate. By providing metrics on the stability of circuits, these techniques can be used to assess the quality and security of quantum code, leading to more reliable quantum products.

\end{abstract}

\section{Introduction}

Quantum computing has recently emerged as a promising alternative to traditional computing, particularly given its optimized processing capabilities for specific optimization problems, which can yield exceptional results \cite{hogg2000quantum,moll2018quantum}. Such challenges are prevalent in various market scenarios, including route optimization for logistics \cite{sales2023adiabatic} or drug discovery \cite{blunt2022perspective}. By leveraging the inherent properties of quantum states, quantum computers can exponentially reduce the complexity of costly algorithmic routines, making them indispensable for industries in the future. However, due to its nascent state, the integration of quantum sub-routines into current software is not a trivial task, especially hindered by the lack of accessibility to quantum computers.

Quantum microservices have presented themselves as a promising solution to these challenges, offering platforms designed for software developers to seamlessly integrate quantum microservices into their final products \cite{garcia2021quantum,alvarado2023quantum}. Additionally, to avoid repeating the mistakes of early software development and to minimize technical debt in the long term, several researchers have advocated for the development of software quality metrics \cite{piattini2022quantum}. These metrics aim to provide developers with additional insights into the quality of their quantum software. Specifically, this paper focuses on circuit stability, a critical quality metric, particularly in terms of software security. 

To measure this property, this paper builds upon the results presented in \cite{alvarado2023improving}. The primary contribution of the paper lies in providing a more suitable circuit processing pipeline to fully take advantage of the potential of traditional machine learning (ML) methods as quantum noise estimators. To achieve this, we replace the complex data extraction from IBM systems with a simpler gate counting strategy, employing ideal and noisy outputs to understand the relationship between circuit structure and resulting discrepancies in the measurements. Our results demonstrate that when circuits are appropriately processed as inputs for the ML model, the outcomes exhibit low error rates. Additionally, we introduce a novel circuit processing technique where gates are transformed into an embedding space, enabling us to treat circuits as images and employ neural networks for noise estimation. In contrast to our prior work, where these models showed suboptimal performance, we achieve competitive accuracy with traditional ML models even in low data regimes.

The structure of this paper is outlined as follows: Subsection \ref{sec:11_quantum_computing} provides a brief introduction to key concepts in the field of quantum computing; Subsection \ref{sec:12_related_work} shows existing results related to quantum noise and estimation algorithms; Section \ref{sec:2_methodology} introduces the methodology employed to adapt ML methods for quantum circuits; Section \ref{sec:3_exp_setup} details the methods and setup employed for conducting the experiments; Section \ref{sec:4_results_and_disc} presents the obtained results with a brief discussion on their advantages and their limitations; and finally, Section \ref{sec:5_conclusions} concludes the paper with a brief summary of the work and outlines future lines of research.  

\subsection{Quantum Computing}
\label{sec:11_quantum_computing}

Quantum computing differs itself from traditional software by employing a probabilistic representation of the states instead of finite and static states. A quantum state is described by a vector in which each coordinate represents the square root of the probability of the system collapsing to its respective state. Formally, a quantum state composed of $n$ qubits is represented by the expression (e.g, a state |5$\rangle$ would be represented as |101$\rangle$).

\begin{equation}
    \sum_{i=0}^{2^n} \alpha_i |i\rangle  = \alpha_1 | \overbrace{0\dots 00}^{n}\hspace{2pt}\rangle + \alpha_2 |0\dots 01\rangle + \dots + \alpha_{2^n} |1\dots 11\rangle
\end{equation}

where $(\alpha_1, \alpha_2, \dots, \alpha_{2^n}) = \mathbf{\alpha} \in \mathbb{R}^{2^n}$ are the coefficients of each state, verifying that $\|\mathbf{\alpha}\|_2 = 1$.

A fundamental property of quantum states is entanglement, which occurs when a quantum state cannot be decomposed into a product of two separate states. This indivisibility implies that traditional computers require exponential time complexity to simulate the transformation of subsequent quantum gates, as each transformation implies multiplying the state by an exponentially large matrix relative to the number of qubits. In contrast, if this property were not fulfilled, traditional computers could exploit state separation to compute the product of the two smaller states separately, thus forfeiting the desired quantum advantage. A well-known example of entanglement is found in Bell states, with one such state expressed as $\frac{1}{\sqrt{2}}\left( |00\rangle+|11\rangle\right)$.


Formally, a NISQ (Noisy Intermediate-Scale Quantum) computer is a system that operates by transforming an initial state $|0\dots0\rangle$ through operations given by quantum gates. The set of possible quantum gates, denoted as $\mathbb{G}$, is constrained based on the specific computer used. Thus, a quantum circuit $\mathbf{C} \in \mathbb{G}^{n\times m}$ can be described as a grid-like structure, where each row represents the path taken by one of the $n$ qubits during execution, and each column contains the gates to be applied to the circuit sequentially. Each cell in this grid structure may be empty, indicating no transformation, or contain a quantum gate, which may involve one (e.g., Hadamard or rotation gates) or multiple qubits (e.g., CNOT or Toffoli gates). According to this formulation, the theoretically expected result when executing the circuit $\mathbf{C}$ is a probability vector in $\mathbb{R}^{2^n}$. However, due to practical limitations, this probability is approximated by a series of consecutive forward passes, involving measurements of the final state and aggregation of the results into a vector representing the distribution of each reported state.

The current state-of-the-art in quantum computers is represented by NISQ devices. However, as their name suggests, these computers are prone to obtaining imprecise results, heavily influenced by various noise sources that undermine result accuracy \cite{resch2021benchmarking}. Two well-known gate-independent noise sources are $T_1$ and $T_2$. The $T_1$ value arises due to state loss in the system, measuring the time for a qubit to transition from a high-energy state $|1\rangle$ to collapse to a $|0\rangle$ state. Similarly, the $T_2$ value measures the dephasing time, representing the duration during which a quantum superposition naturally collapses to either $|0\rangle$ or $|1\rangle$. Another significant noise source arises from imperfect gates, where weak calibration can cause states to deviate from the expected values, potentially resulting in drastically different outcomes \cite{bravyi2018correcting}.

\subsection{Related Work}
\label{sec:12_related_work}





Quantum noise is a well-documented phenomenon that can significantly affect the accuracy of circuit results if not properly addressed \cite{clerck2010,resch2021benchmarking}. As mentioned in the previous section, the origins of this noise are multifactorial, leading to non-trivial heuristics for estimating whether a circuit will produce a confident enough output distribution compared to the theoretical benchmark \cite{resch2021benchmarking,bravyi2018correcting}. However, this quality metric is of critical importance, as executions in low robustness environments can yield erroneous results.

To tackle this issue, two main directions are being studied. The first branch focuses on the hardware aspects that give rise to noise, aiming to enhance the stability and robustness of quantum computers while minimizing the misalignment introduced by quantum gates. The second branch, represented by error correction codes, draws on traditional techniques to prevent information loss or modification from impacting the entire system \cite{knill1997theory}. These methods have already demonstrated their ability to deliver more accurate outputs at the expense of requiring additional qubits to serve as redundancy for computations \cite{calderbank1996good}. However, both approaches primarily concentrate on reducing noise during execution. Neither technique adequately addresses the challenge of determining whether a circuit will perform poorly under a specific noise model.

In order to solidify the quality of quantum circuits for production-ready software, a novel branch of noise estimation has recently emerged. This branch, referred to as quantum noise estimation algorithms, focuses on developing algorithms that can estimate the deviation from theoretical to practical quantum states within a specific quantum system without the need for either simulation or execution on a real quantum system. This estimation involves the development of a function $\mathcal{N} : \mathbf{C} \rightarrow \mathbb{R}$, which maps a circuit to a scalar value, quantifying the distance between real and practical errors.

Currently, literature on this topic is scarce, however, two approaches have already been proposed. The first method was introduced by Aseguinolaza et al., advocating for the use of an ad-hoc formula to measure the fidelity output of quantum circuits. The authors analyzed the noise effects in several well-known quantum algorithms, demonstrating how their metric can determine result robustness \cite{aseguinolaza2023error}. A second approach was presented by Alvarado et al., where they utilized a linear method to measure the estimated difference between theoretical and actual outputs, based on IBM models \cite{alvarado2023improving}. This approach aims to provide a quality metric for incorporation into a quantum micro-service pipeline, serving as a security measure for developers.

\section{Methodology}
\label{sec:2_methodology}

This section describes the process employed to clean and process the quantum circuits to allow machine learning methods to estimate the noise. This section starts by introducing the process employed to generate the training data. Then, it details the transformation process employing for using the data for traditional methods focused on tabular data. Finally, it details the process employed to transform the circuits into a well-suited data shape for conventional neural networks. 

\textbf{Dataset Generation}\hspace{0.3cm} To create a large and representative dataset of distinct quantum circuits, we generated 35,000 random quantum circuits using the quantum gates available on IBM machines. These 35,000 samples were divided into seven chunks of 5,000 circuits each, with each chunk containing circuits of varying qubit counts, ranging from 4 to 10 qubits. The depth of each circuit was selected from a uniform distribution, spanning from short circuits with only 2 columns to deeper instances with up to 50 columns.

\textbf{Tabular Data Adaptation}\hspace{0.3cm} Similar to the approach employed in \cite{alvarado2023improving}, we utilize a function $N_{\textup{Emb}}^T$, defined as the mapping of the circuit to its gate count, which provides a more suitable space for tabular methods. While we collected gate errors from IBM machines, we opted not to incorporate them in this work. This decision stems from the increased data complexity associated with individual errors across multiple gates and qubits. Therefore, we hypothesize that by limiting the amount of information fed to the circuits, models can improve their learning dynamics, extracting simpler yet more robust relationships between circuit composition information and obtained noise measurements.

\textbf{Neural Network Data Adaptation}\hspace{0.3cm} In the aforementioned work, we also employed neural networks to estimate noise; however, the experiments resulted in the neural networks being unable to learn. We attributed this outcome to the tabular nature of the data, which is known to be unsuitable for these approaches. To address this limitation, we propose transforming quantum circuits into low-dimensional image-like samples. Given the grid-like structure of a circuit $\mathcal{C}$ with a depth $D$ and $Q$ qubits, we define a function $\mathcal{N}_{Emb}^I : \mathbb{G}^{D \times Q} \rightarrow \mathbb{R}^{E \times D \times Q}$ to project the given circuit into a low-dimensional representation using an embedding dimension of $E$.

This projection function $\mathcal{N}_{Emb}^I$ consists of two sequential transformations. The first transformation maps the circuit from the gate space $\mathbb{G}$ to the space of natural numbers $\mathbb{N}$, simplifying each circuit into a more manageable space. This transformation can be achieved by assigning each unique gate to a consecutive integer, thus establishing a mapping of distinct gates to consecutive integers. Given the poor performance observed by neural networks when operating in such spaces, we apply an additional learnable embedding to this set, enabling each gate to be represented by a vector in $\mathbb{R}^{E}$. With this formulation, neural networks should be better equipped to leverage circuit information for optimal noise estimation.

\section{Experimental Setup}
\label{sec:3_exp_setup}

IBM offers a variety of quantum computers for academic use, each featuring a distinct noise model. To evaluate the quantum computer's impact on our approach, we selected two backends for experimentation: IBM Cairo and IBM Hanoi, both capable of handling circuits with up to 27 qubits. Due to the substantial queue required to compute measurements on a dataset as extensive as the one proposed in this work using real machines, we chose to simulate the results. We employed the Qiskit Aer simulator \cite{Qiskit}, conducting 1000 shots for each circuit in both noisy and ideal executions. The distance between these two outputs has been measured using the cosine distance.

Due to the inherent high costs of computing a large set of circuits, we chose to evaluate the robustness of these methods against different training sizes. We conducted experiments using two training split sizes: 1\% and 80\%. Each training split was obtained by randomly sampling the specified size of random samples from the aforementioned 35,000 generated circuits\footnote{Dataset repository: \url{https://github.com/Jongarde/Qserv_dataset}}. No additional data augmentation techniques were employed during training. Two metrics were used to compute the error between the estimated noise and the actual computed noise, namely, the Mean Absolute Error (MAE) and the Root Mean Squared Error (RMSE). 

To compare our results, we employ tabular-based methods under $\mathcal{N}_{Emb}^T$, and neural networks under the $\mathcal{N}_{Emb}^I$ projection. The chosen tabular methods include Linear Regression, Histogram-based Gradient Boosting \cite{shi2022quantized}, and XGBoost \cite{chen2016xgboost}. For neural network models, we selected a dense neural network and a convolutional neural network. No hyperparameter tuning operation was conducted for any algorithm.

Both neural networks have approximately 130,000 parameters. The dense neural network comprises 3 layers with 80, 40, and 1 neurons, respectively. The convolutional neural network consists of 2 convolutional layers, each with 16 channels, followed by 3 dense layers with the same structure as the aforementioned neural network. Both networks were trained using the Mean Square Error loss with the Adam Optimizer for 100 epochs. The learning rate was set to 0.001, halving every 5 epochs during the first 50 epochs.

\begin{figure}[t]
  \centering
  \includegraphics[width=\textwidth]{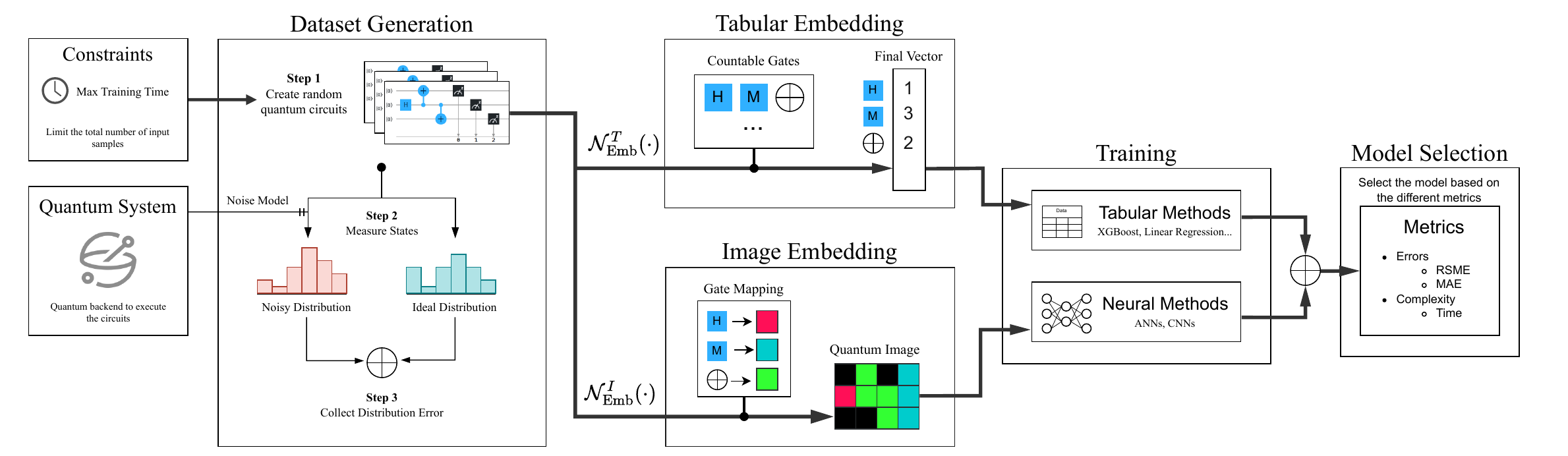}
  \caption{Schema detailing the process followed to obtain the data and train the different models.}\label{fig:pipeline}
\end{figure}

\section{Results \& Discussions}
\label{sec:4_results_and_disc}

The results of the error for all training methods are presented in Table \ref{tab:result}. Under these results, our initial hypothesis on the accuracy of these methods in estimating noise is verified. It comes as no surprise to observe that linear regression, which closely resembles our prior approach, yields the least optimal results. This method only captures linear dependencies between the number of gates and the error, neglecting further relationships within the circuits. This defect is particularly clear in the data related to IBM Hanoi, where the model has an almost double error rate. In contrast, all other methods attain competitive results, with XGBoost yielding the most precise outcomes.

\begin{table}[h]
    \centering
    \caption{MAE and RSME of the ideal and noisy outputs of the quantum circuits under all ML methods. Each cell contains the error when employing a train split size of 1\% (left-side of the arrow), and the error when employing an 80\% split size (right-side of the arrow).}
    \label{tab:result}
    \resizebox{\textwidth}{!}{
    \begin{tabular}{lcccccccc}
        \toprule[1.3pt]
        \midrule[0.2pt]
        & & \multicolumn{3}{c}{\textbf{IBM Cairo}} & &\multicolumn{3}{c}{\textbf{IBM Hanoi}} \\
        \textbf{Algorithm} &  & {\footnotesize MAE} && {\footnotesize RSME} && {\footnotesize MAE} && {\footnotesize RSME} \\
        \midrule
        
        XGboost & \hspace{0.2cm}$\text{ }$& 0.082 $\rightarrow$ 0.073 &\hspace{0.2cm}$\text{ }$& 0.120 $\rightarrow$ 0.106 &\hspace{0.4cm}$\text{ }$& 0.051 $\rightarrow$ 0.042&\hspace{0.2cm}$\text{ }$&0.083 $\rightarrow$ 0.073\\
        
        Hist Gradient Boost & \hspace{0.2cm}$\text{ }$& 0.081 $\rightarrow$ 0.070 &\hspace{0.2cm}$\text{ }$& 0.117 $\rightarrow$ 0.102 &\hspace{0.4cm}$\text{ }$& 0.050 $\rightarrow$ 0.039&\hspace{0.2cm}$\text{ }$&0.081 $\rightarrow$ 0.069\\
        
        Linear Regression& \hspace{0.2cm}$\text{ }$& 0.093 $\rightarrow$ 0.086 &\hspace{0.2cm}$\text{ }$& 0.122 $\rightarrow$ 0.115 &\hspace{0.4cm}$\text{ }$& 0.090 $\rightarrow$ 0.083&\hspace{0.2cm}$\text{ }$&0.116 $\rightarrow$ 0.107\\
        
        Dense NN & \hspace{0.2cm}$\text{ }$& 0.081 $\rightarrow$ 0.074 &\hspace{0.2cm}$\text{ }$& 0.111 $\rightarrow$ 0.106 &\hspace{0.4cm}$\text{ }$& 0.058 $\rightarrow$ 0.043 &\hspace{0.2cm}$\text{ }$& 0.088 $\rightarrow$ 0.071 \\

        Convolutional NN & \hspace{0.2cm}$\text{ }$& 0.086 $\rightarrow$  0.073&\hspace{0.2cm}$\text{ }$& 0.117 $\rightarrow$ 0.110 &\hspace{0.4cm}$\text{ }$& 0.066 $\rightarrow$ 0.042 &\hspace{0.2cm}$\text{ }$& 0.096 $\rightarrow$ 0.074 \\
        \midrule[0.2pt]
        \bottomrule[1.1pt]
    \end{tabular}
    }
\end{table}

Nevertheless, a clear pattern can be observed when analyzing the learned heuristics of tabular-based ML methods, as most of them heavily rely on gates related to the total qubit count of the circuit, indicating a strong correlation between this value and the total system's noise. While this effect was expected, as higher qubit counts imply greater entanglement and thus more decoherence-related noise, the impact of this variable surpasses initial hypotheses. To further analyze this effect, we examine the effect of removing gates related to the qubit count (e.g. reset or measure), and presented the results in Table \ref{tab:result_no_gate}. A clear downtrend in all algorithms is appreciated. This effect is particularly pronounced when looking into the results of IBM Hanoi, where the amount of qubits serves as the leading noise indicator. Interestingly, under this framework, Gradient Boosting methods fail to establish previous patterns relating the qubit count to other gates, causing Linear Regression to transition from the least accurate among all methods to the one with the highest accuracy.

Moreover, our results also prove how employing non-tabular inputs for neural network-based approaches can yield competitive results. While these models previously underperformed when incorporating additional variables from the IBM noise models in \cite{alvarado2023improving}, the reduced amount of extra information has led to increased accuracy. The results show competitiveness against gradient boosting methods, especially in both low and high data regimes. Due to the data processing pipeline, where circuits are transformed into image representations, the information regarding the number of qubits cannot be hidden and is therefore not included in Table \ref{tab:result_no_gate}.

\begin{table}[h]
    \centering
    \caption{MAE and RSME of the ideal and noisy outputs of the quantum circuits under tabular methods with \texttt{reset} and \texttt{measure} gates removed. Each cell contains the error when employing a train split size of 1\% (left-side of the arrow), and the error when employing an 80\% split size (right-side of the arrow).}
    \label{tab:result_no_gate}
    \resizebox{\textwidth}{!}{
    \begin{tabular}{lcccccccc}
        \toprule[1.3pt]
        \midrule[0.2pt]
        & & \multicolumn{3}{c}{\textbf{IBM Cairo}} & &\multicolumn{3}{c}{\textbf{IBM Hanoi}} \\
        \textbf{Algorithm} &  & {\footnotesize MAE} && {\footnotesize RSME} && {\footnotesize MAE} && {\footnotesize RSME} \\
        \midrule
        
        XGboost & \hspace{0.2cm}$\text{ }$& 0.102 $\rightarrow$ 0.094 &\hspace{0.2cm}$\text{ }$& 0.132 $\rightarrow$ 0.121 &\hspace{0.4cm}$\text{ }$& 0.138 $\rightarrow$ 0.113&\hspace{0.2cm}$\text{ }$&0.178 $\rightarrow$ 0.154\\
        
        Hist Gradient Boost & \hspace{0.2cm}$\text{ }$& 0.098 $\rightarrow$ 0.091 &\hspace{0.2cm}$\text{ }$& 0.127 $\rightarrow$ 0.117 &\hspace{0.4cm}$\text{ }$& 0.130 $\rightarrow$ 0.110&\hspace{0.2cm}$\text{ }$&0.168 $\rightarrow$ 0.149\\
        
        Linear Regression& \hspace{0.2cm}$\text{ }$& 0.095 $\rightarrow$ 0.092 &\hspace{0.2cm}$\text{ }$& 0.122 $\rightarrow$ 0.119 &\hspace{0.4cm}$\text{ }$& 0.124 $\rightarrow$ 0.111&\hspace{0.2cm}$\text{ }$&0.161 $\rightarrow$ 0.150\\
        
        \midrule[0.2pt]
        \bottomrule[1.1pt]
    \end{tabular}
    }
\end{table}

One crucial characteristic of a reliable noise estimator for production-ready software is the speed of computing this metric. Given the simple design of the selected algorithms and the small size of the parameters in the neural networks, our models can perform extremely quickly, even computing hundreds or thousands of instances per second. These timing results are presented in Table \ref{tab:result_time}, where we measure the time taken to obtain the results of training and testing in each respective set of circuits. The speed advantages are clearly evident in traditional machine learning methods, as they can process up to 10,000 instances per second. This effect is particularly noticeable when using linear regression methods. However, when seeking a more robust method, XGBoost can achieve competitive accuracies while maintaining high prediction speeds. In comparison, these models achieve much faster training speeds than neural networks, given the costly training loop required to update these models. Moreover, between standard neural networks and convolutional neural networks, convolutional require twice as much computation time for both training and inference. 

\begin{table}[h]
    \centering
    \caption{Time in seconds of execution of the training and prediction process under each method. We showcase the results under the $1\%$ and the $80\%$ training splits.}
    \label{tab:result_time}
    \resizebox{\textwidth}{!}{
    \begin{tabular}{lcS[table-format=2.3]S[table-format=2.3]S[table-format=2.3]S[table-format=2.3]cccc}
        \toprule[1.3pt]
        \midrule[0.2pt]
        & & \multicolumn{3}{c}{\textbf{Training}} & &\multicolumn{3}{c}{\textbf{Prediction}} \\
        \textbf{Algorithm} &  & {\footnotesize 1\%} && {\footnotesize 80\%} && {\footnotesize 1\%} && {\footnotesize 80\%} \\
        \midrule
        
        XGboost & \hspace{0.5cm}$\text{ }$& 0.086 &\hspace{0.2cm}$\text{ }$& 2.940 &\hspace{0.4cm}$\text{ }$& 0.068&\hspace{0.2cm}$\text{ }$&0.062\\
        
        Hist Gradient Boost & \hspace{0.5cm}$\text{ }$& 0.225 &\hspace{0.2cm}$\text{ }$& 1.437 &\hspace{0.4cm}$\text{ }$& 0.242&\hspace{0.2cm}$\text{ }$&0.088\\
        
        Linear Regression& \hspace{0.5cm}$\text{ }$& 0.016 &\hspace{0.8cm}$\text{ }$& 0.258 &\hspace{1cm}$\text{ }$& 0.015&\hspace{0.8cm}$\text{ }$&0.016\\
        
        Dense NN & \hspace{0.5cm}$\text{ }$& 5.417 &\hspace{0.2cm}$\text{ }$& 413.916 &\hspace{0.4cm}$\text{ }$& 2.085 &\hspace{0.2cm}$\text{ }$& 0.428 \\

        Convolutional NN & \hspace{2.2cm}$\text{ }$& 10.720 &\hspace{0.75cm}$\text{ }$& 729.841 &\hspace{2cm}$\text{ }$& 3.684 &\hspace{0.75cm}$\text{ }$& 0.843 \\
        \midrule[0.2pt]
        \bottomrule[1.1pt]
    \end{tabular}
    }
    
\end{table}

\section{Conclusion}
\label{sec:5_conclusions}


In this study, we investigate the reliability of machine learning methods in obtaining accurate noise estimations in quantum circuits. To accomplish this, we created a large dataset of circuits with varying numbers of qubits, gates, and depth. Subsequently, we computed the distances between noisy and ideal output distributions, leveraging two distinct noise models extracted from IBM backends. Through a well-suited data processing pipeline that solely relies on the gate counts of the circuits, we trained three well-known ML models to estimate this noise metric. Additionally, recognizing the limited capability of neural networks to handle tabular data, we expanded the pipeline to obtain image representations of the circuits, which are particularly good suit for convolutional neural networks. Our results demonstrate the robustness of these methods in providing accurate heuristics for estimating the fidelity of the circuit's output. Moreover, given the speed offered by these algorithms, these techniques are particularly appropriate for robust production systems requiring high throughput. We envision a future line of focused on integrating the concepts developed in this paper into a production-ready tool for analyzing quantum software and quantum code.

\section*{Acknowledgements}

The authors would like to acknowledge the partial financial support by Ministry of Science (project QSERV-UD, PID2021-124054OB-C33), and also to the Basque Government (projects TRUSTIND - KK-2020/00054, and REMEDY - KK-2021/00091). Additionally, the authors wish to acknowledge the selfless support from IBM, who generously provided their quantum computing equipment for the project. Finally, it is important to also express gratitude for the support and drive that the regional government of Bizkaia is providing in all matters related to the development of quantum technologies as a driving force for progress of the Society of this historic territory.

\renewcommand{\contentsname}{References}
\renewcommand{\refname}{References}
\bibliographystyle{IEEEtran}
\bibliography{bibliography}

\end{document}